\documentclass[prl,aps,twocolumn,groupedaddress,showpacs]{revtex4}

\usepackage{graphicx}
\begin{document}

\title{ Bragg diffraction of microcavity polaritons by a surface
acoustic wave }

\author{Kikuo Cho and Kazunori Okumoto}

\affiliation{Graduate School of Engineering Science, Osaka
University, 1-3 Machikaneyama, Toyonaka 560-8531, Japan}

\author{N.~I. Nikolaev and A.~L. Ivanov}

\affiliation{Department of Physics and Astronomy, Cardiff
University, Cardiff CF24 3YB, United Kingdom}

\date{\today}

\begin{abstract}
Bragg scattering of polaritons by a coherent acoustic wave is
mediated and strongly enhanced by the relevant exciton states
resonant with the acoustic and optic fields simultaneously. In
this case, in sharp contrast with conventional acousto-optics, the
resonantly enhanced Bragg spectra reveal the multiple orders of
diffracted light, i.e., a Brillouin band structure of
parametrically driven polaritons can be directly visualized. We
analyze the above scheme for polaritons in (GaAs) semiconductor
microcavities driven by a surface acoustic wave (SAW) and show
that for realistic values of the SAW, $\nu_{\rm SAW} = 1$\,GHz and
$I_{\rm ac} \lesssim 100$\,W/cm$^2$, the main acoustically-induced
band gap in the polariton spectrum can be as large as $\Delta^{\rm
MC}_{\rm ac} \simeq 0.6$\,meV, and the Bragg replicas up to $n=3$
can be observed.
\end{abstract}


\maketitle

Back in 1933, L\'eon Brillouin, a founder of acousto-optics,
described for the first time non-perturbative solutions of the
Maxwell wave equation for the dielectric constant $\varepsilon$
harmonically modulated by an acoustic wave $\{ {\bf k},\Omega_{\bf
k}^{\rm ac}\}$, i.e., for $\varepsilon = \varepsilon_b + \delta
\varepsilon_b \cos(\Omega_{\bf k}^{\rm ac}t - {\bf k}{\bf r})$
\cite{Brillouin33}. Formally, the eigenvalues and eigenstates of
the above equation are given in terms of the Mathieu functions and
yield an infinite sequence of the alternating allowed and
forbidden energy bands in momentum space. While the energy
structure is akin to the electron energy bands, which appear in
the presence of a periodic (atomic) potential, a very particular
feature of the {\it harmonic} potential is that the forbidden
energy bands, i.e., the acoustically-induced energy gaps
$\Delta_{\rm ac}^{(n)}$ ($n=1,2,..$) in the photon spectrum,
extremely effectively close up with the increasing energy number
$n$: $\Delta_{\rm ac}^{(n)} \propto (\delta \varepsilon_b)^n$.
This is an inherent property of the Mathieu wavefunctions
\cite{Blanch65}. The acousto-optic science deals with $\delta
\varepsilon_b = 4 \pi \chi^{(2)}_{\gamma-{\rm ac}} I_{\rm
ac}^{1/2}$, where $I_{\rm ac}$ is the intensity of the acoustic
wave and $\chi^{(2)}_{\gamma-{\rm ac}}$ is a second-order
nonresonant acousto-optic susceptibility \cite{Wilson83,Korpel97}.
The nonresonant acousto-optic nonlinearities are small, so that
usually $\delta \varepsilon_b \sim 10^{-4}\!- 10^{-3}$ (for GaAs,
e.g., $\delta \varepsilon_b \simeq 1.6\!\times\!10^{-3}$ for
$I_{\rm ac} = 100$\,W/cm$^2$). As a result, the conventional
acousto-optics, both fundamental and applied, attributes the Bragg
diffraction of the light field by a sound wave to the one-phonon
transition only, i.e., to the main acoustically-induced gap $n=1$
\cite{Wilson83,Korpel97,Brillouin22,Chu76}.

The nonresonant acousto-optic nonlinearities, which give rise to
the acoustically-induced diffraction grating, is due to the
photoelastic effect. In sharp contrast with this mechanism, the
{\it resonant acousto-optic effect} \cite{Ivanov01,Ivanov03} for
bulk and microcavity (MC) polaritons deals with a quantum
diffraction of optically-induced, virtual excitons by the coherent
acoustic wave. In this case the interaction of two ``matter''
waves, the excitonic polarization and the acoustic field, is much
more effective than the nonresonant photoelastic coupling of
photons with acoustic phonons. In the meantime the photon
component of optically-dressed excitons, i.e., of polaritons, can
be large enough to ensure an efficient resonant conversion
``photon $\leftrightarrow$ exciton''. This explains the origin of
the resonant, exciton-mediated acousto-optic nonlinearity, $\delta
\varepsilon^{\rm x}_{\gamma-{\rm ac}} \gg \delta \varepsilon_b$.
For GaAs-based structures at low temperatures $\delta
\varepsilon^{\rm x}_{\gamma-{\rm ac}}$ can be as large as $\delta
\varepsilon^{\rm x}_{\gamma-{\rm ac}} \sim 0.1$. Thus the resonant
acousto-optics of polaritons deals with the microscopic mechanism
fundamentally different from the photoelastic effect.

In this Letter we propose and calculate the acousto-optic Bragg
scattering of SAW-driven polaritons. The MC polariton eigenstates
in GaAs $\lambda$-cavities \cite{Skolnick98,Baumberg02,Kavokin03}
are relevant to the proposed Bragg spectroscopy, due to
compatibility of GaAs microcavities with the SAW technique
\cite{Campbell89} and due to a possibility to realize
one-dimensional geometry for the resonant, SAW-mediated
interaction of two counter-propagating MC polaritons by using
$\nu_{\rm SAW} \lesssim 3$\,GHz \cite{Ivanov03}. As we demonstrate
below, the Bragg spectra of MC polaritons are resonantly enhanced
by the quantum well (QW) exciton state and consist of the
well-developed multiple replicas with a robust structure.

For the acousto-optic spectroscopy of polaritons, both resonant
interactions, the exciton-photon coupling and the interaction of
excitons with the acoustic pump wave, should be treated
non-perturbatively (strong coupling regime) and on an equal basis.
Being applied to the Bragg scattering of MC polaritons, the
acousto-optic polariton macroscopic equations \cite{Ivanov01} are
given by

\begin{equation}
 \left[ { \partial^2 \over \partial x^2 } + { \partial^2 \over
\partial z^2 } - { \varepsilon_b(z) \over c^2 }
{ \partial^2 \over \partial t^2 } \right] E = { 4\pi \over c^2 }{
\partial^2 \over \partial t^2 }P + J_{\rm ext} \, ,
\label{macroE}
\end{equation}

\begin{eqnarray}
\nonumber \bigg[ { \partial^2 \over
\partial t^2 } + 2 \gamma_{\rm x} {
\partial \over \partial t } + \omega_{\rm T}^2 - { \hbar \omega_{\rm T}
\over M_{\rm x} } { \partial^2 \over \partial x^2 } - \\
- 4m_k^{\rm x}\, \omega_{\rm T} \cos( \Omega^{\rm ac}_k t - kx)
\bigg] P = \Omega_{{\rm x}-\gamma}^2 E \, , \label{macroP}
\end{eqnarray}

where $E$ and $P$ are the light field and excitonic polarization,
respectively, $J_{\rm ext}$ is a source of the external optical
wave necessary for the Bragg scattering problem, $\hbar
\omega_{\rm T}$ is the energy position of the exciton line,
$M_{\rm x}$ is the in-plane translational mass of QW excitons,
$\Omega_{{\rm x}-\gamma}$ is the matrix element of the QW exciton
-- MC photon interaction, and $2\gamma_{\rm x} = 1/T_1 = 2/T_2$ is
the rate of incoherent scattering of QW excitons. We assume a
Cartesian coordinate system with the in-plane $x$-axis along the
SAW wavevector ${\bf k}$, the $z$-axis along the MC growth
direction, and the exciton and light fields linearly polarized
along the in-plane $y$-axis. In this case the SAW, which has both
transverse and longitudinal displacement components, $u_x$ and
$u_z$, is elliptically-polarized in the $x$-$z$ plane (the
sagittal plane).

The SAW-induced modulation of the excitonic polarization is
characterized by the coupling coefficient $m_k^{\rm x} = I_{\rm
ac}^{1/2} [|u_x|/(|u_x|^2 + |u_z|^2)^{1/2}] |m_{\rm x-ac}^{\rm DP}
+ i \gamma_{\rm saw} m_{\rm x-ac}^{\rm PE}|$, where $m_{\rm
x-ac}^{\rm DP}$ and $m_{\rm x-ac}^{\rm PE}$ are the deformation
potential (DP) and piezoelectric (PE) matrix elements of exciton
-- SAW interaction, respectively, and $\gamma_{\rm saw} =
|u_z|/|u_x| \exp(i \delta_{\rm saw})$. The phase shift
$\delta_{\rm saw}$ between $u_x$ and $u_z$ is nearly $\pi/2$
\cite{Campbell89}, so that both interaction channels interfere
constructively. However, because $m_{\rm x-ac}^{\rm DP} \gg m_{\rm
x-ac}^{\rm PE}$ (for $\nu_{\rm SAW} \lesssim 10$\,GHz) and
$|u_z|/|u_x| \simeq 1.4$ \cite{Simon96}, the DP mechanism with
$m_{\rm x-ac}^{\rm DP} = D_{\rm x}/(2 \hbar^2 \rho v_{\rm
s}^3)^{1/2}$ is strongly dominant over the PE one. Here, $\rho$ is
the crystal density, $v_{\rm s}$ is the SAW velocity, and $D_{\rm
x}$ is the exciton -- LA-phonon DP. The above hierarchy of the
interaction mechanisms  is due to the charge neutrality of
excitons: In contrast, the electron -- SAW interaction is
determined by the PE coupling \cite{Simon96}.

The $z$-dependent background dielectric function
$\varepsilon_b(z)$ on the left-hand-side (l.h.s.) of
Eq.\,(\ref{macroE}) refers to a stack of layers which form the MC
structure we analyze, i.e., a GaAs $\lambda$-cavity with an
embedded InGaAs QW, symmetrically sandwiched in between two
identical AlGaAs/AlAs Bragg reflectors. The Bragg reflectors give
rise to the transverse optical confinement and, in the meantime,
ensure an optical coupling of the external light wave with the
in-plane polariton quasi-eigenstates. At first, however, in order
to visualize the quasi-energy spectrum of SAW-driven MC
polaritons, we assume $100\,\%$ optical confinement (i.e., replace
the Bragg reflectors by perfect mirrors), no external light source
[i.e., $J_{\rm ext}=0$ on the r.h.s. of Eq.\,(\ref{macroE})], and
no damping of the exciton states [i.e., $\gamma_{\rm x} = 0$ on
the l.h.s. of Eq.\,(\ref{macroP})]. In this case the polariton
spectrum is characterized by the MC Rabi frequency $\Omega^{\rm
MC}_{\rm x} = 2 (\pi / \varepsilon_b)^{1/2} \Omega_{{\rm
x}-\gamma}$. In Fig.\,1 we plot the quasi-energy spectrum of
lower-branch (LB) polaritons in a zero detuning MC with
$\Omega^{\rm MC}_{\rm x} = 3.7$\,meV, driven by the SAW of
$\nu_{\rm SAW} = 1$\,GHz and $I_{\rm ac} = 10$\,W/cm$^2$.

\begin{figure}
\includegraphics*[width=8cm]{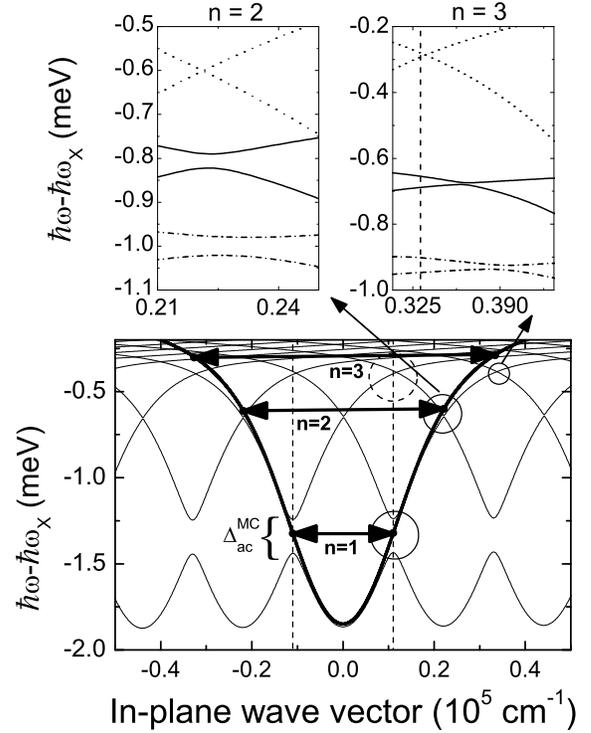}
\caption{ The acoustically induced quasi-energy spectrum of
polaritons in a GaAs $\lambda$-microcavity driven by the SAW of
$\nu_{\rm SAW} = 1$\,GHz, $k = 2.2\!\times\!10^4$\,cm$^{-1}$,
$I_{\rm ac} = 10$\,W/cm$^2$, $v_{\rm s} = 2.9 \times 10^5$\,cm/s,
and $D_{\rm x} = 12$\,eV (thin solid lines). The initial,
acoustically-unperturbed lower dispersion branch of MC polaritons
is shown by the bold solid line. The MC Rabi energy is
$\hbar\Omega_{\rm x}^{\rm MC} = 3.7$\,meV. The magnified energy
bands relevant to the resonant $n = 2$ and 3 transitions are
plotted in the l.h.s. and r.h.s. upper insets, respectively, for
$I_{\rm ac} = 0$ (dotted lines), $I_{\rm ac} = 50$\,W/cm$^2$
(solid lines), and $I_{\rm ac} = 100$\,W/cm$^2$ (dash-dotted
lines). The vertical dashed lines show the boundaries of the
acoustically-induced first Brillouin zone. }
\label{fig1}
\end{figure}

The quasi-energy spectrum, which is calculated by the method
developed in Ref.\,\cite{Ivanov01} for bulk polaritons and cannot
be interpreted in terms of the Mathieu functions, show the
acoustically-induced band gaps $\Delta_{\rm ac}^{\rm MC (n)}
\propto I_{\rm ac}^{n/2}$, due to the $n=1,2,...$ --
phonon-assisted transitions. Because the energy of SAW phonons is
very small, $h \nu_{\rm SAW} \simeq 4.1\,\mu$eV only, the
acoustically-induced spectral gaps refer to the in-plane
counter-propagating MC polariton states $\{nk/2,\omega^{\rm
MC}_{\rm LB}(nk/2) \}$ and $\{ -nk/2,\omega^{\rm MC}_{\rm
LB}(nk/2) \}$ (see the nearly horizontal arrows in Fig.\,1), where
$\omega^{\rm MC}_{\rm LB}(p_{\|})$ with $p_{\|}= \pm n k/2$ is the
LB solution of the dispersion equation $c^2 p_{\|}^2/\varepsilon_b
+ \omega_{\rm T}^2 - \omega^2= (\omega \Omega^{\rm MC}_{\rm
x})^2/(\omega_{\rm T}^2 + \hbar \omega_{\rm T} p_{\|}^2/M_{\rm x}
- \omega^2)$. The main acoustically-induced band gap in the LB
spectrum is given by $\Delta_{\rm ac}^{\rm MC} \equiv \Delta_{\rm
ac}^{\rm MC (n=1)} = 2 |m_k^{\rm x}| \varphi^{\rm MC}(k/2)$, where
$\varphi^{\rm MC}(k/2) = (\Omega^{\rm MC}_{\rm x})^2/\{
(\Omega^{\rm MC}_{\rm x})^2 + 4 [\omega_{\rm T} - \omega^{\rm
MC}_{\rm LB}(k/2)]^2\}$ is the excitonic component of the
polariton states $p_{\|}  = \pm k/2$ resonantly coupled via
one-SAW-phonon transition. Even for modest SAW intensities the
main gap is large, so that $\Delta_{\rm ac}^{\rm MC} \gg h
\nu_{\rm SAW}$. For example, $I_{\rm ac} = 100$\,W/cm$^2$ yields
$\Delta_{\rm ac}^{\rm MC}/h \simeq 130$\,GHz. Note, that the
conventional acousto-optics deals with the acoustically-induced
stop gap $\Delta_{\gamma-{\rm ac}}$ less than the SAW frequency
(usually, $\Delta_{\gamma-{\rm ac}} \lesssim 100$\,MHz)
\cite{Korpel97,Campbell89}. Furthermore, the acoustic band gaps
associated with the two- and three-phonon transitions in the MC
polariton spectrum are also well developed, so that $\Delta_{\rm
ac}^{\rm MC (n=1)} : \Delta_{\rm ac}^{\rm MC (n=2)} : \Delta_{\rm
ac}^{\rm MC (n=3)} = 1:0.08:0.03$ for $I_{\rm ac} = 100$\,W/cm$^2$
(see the insets of Fig.\,1). This is in a sharp contrast with the
traditional acousto-optic schemes of the Bragg scattering. All the
above features of the quasi-energy spectrum of SAW-driven MC
polaritons are due to the resonant, exciton-mediated acousto-optic
susceptibility.

Due to the periodicity of the acoustic wave, the quasi-energy
spectrum can also be interpreted in terms of an {\it extended}
Brillouin zone, with the band boundaries at $p_{\|} \simeq \pm
nk/2$. The first acoustically-induced Brillouin zone is shown in
Fig.\,1 by the vertical dashed lines. The energy band boundaries,
where the spectral gaps arise and develop with increasing $I_{\rm
ac}$, can be probed in Bragg scattering, by changing the incidence
angle $\alpha$ of the external optical wave. At this point we
return to Eqs.\,(\ref{macroE})-(\ref{macroP}) applied to a
realistic MC structure with the Bragg reflectors consisting of 34
alternating AlAs and Al$_{0.13}$Ga$_{0.87}$As $\lambda/4$-layers.
The MC polariton quasi-eigenstates can now be excited by the
external light field, $E_{\rm inc} = E_{\rm inc}^{(0)}e^{-i\omega
t}$ [due to the $J_{\rm ext}$-term on the r.h.s. of
Eq.\,(\ref{macroE})], and can decay or scatter into the external
electromagnetic modes.

In order to calculate the SAW-induced Bragg diffraction of
optically excited MC polaritons, we use the Green function
technique developed in Ref.\,\cite{Cho03}. Because the in-plane
wavevector $p_{\|}$ is conserved in the propagation of the light
field through an optically transparent planar structure, one
defines the MC photon Green function $g(z,z';\omega)$ by the
equation:
\begin{equation}
{ d^2 \over dz^2 }\,g(z,z';\omega) + \kappa^2 g(z,z';\omega) = -
4\pi \delta(z-z')\,,
\label{green}
\end{equation}
where $\kappa^2 = (\omega/c)^2 \varepsilon_b(z) -
p_{\parallel}^2$. The function $g(z,z';\omega)$, which satisfies
the Maxwellian boundary conditions, is evaluated numerically with
the use of $\varepsilon_b(z)$ relevant to the MC structure we
analyze. As a next step, we substitute in
Eqs.\,(\ref{macroE})-(\ref{macroP}) a Fourier expansion of the $E$
and $P$ fields in terms of the in-plane quasi-wavevector $p_{\|} +
\ell k$ and quasi-frequency $\omega + 2 \pi \ell \nu_{\rm SAW}$,
and evaluate the SAW-mediated acousto-optic susceptibility matrix
$\chi_{\ell,\ell'}$: $P_{\ell} = \sum_{\ell'} \chi_{\ell,\ell'}
E_{\ell'}$ ($\ell = 0,\pm1,..$, i.e., $\ell = \pm n$ corresponds
to the $n$-phonon transition). The acoustically-induced
polarization harmonics $P_{\rm \ell}$ give rise to the Bragg
signal:
\begin{equation}
E_{\ell} =  E_{\ell}^{(0)} + q_{\ell}^2 \int \mbox{d}z'
g(z,z',\omega_{\ell}) \sum_{\ell'} \chi_{\ell,\ell'} E_{\ell'}(z') \,,
\label{signal}
\end{equation}
where $q_{\ell}^2 = (\omega + 2 \pi \ell \nu_{\rm SAW})^2/c^2$,
$E_{\ell} = E_{\ell}(z)$ is the signal light field, and
$E^{(0)}_{\ell} = E^{(0)}_{\ell}(z)$ is the incoming light field
induced by $E_{\rm inc}$. The field $E^{(0)}_{\ell}(z)$ is
calculated by using the photon Green function defined by
Eq.\,(\ref{green}). The integration on the r.h.s. of
Eq.\,(\ref{signal}) is over the QW thickness, so that one can
approximate $E_{\ell'}(z')$ by $E_{\ell'}(z'\!\!=\!\!0)$.
The latter amplitude is evaluated from Eq.\,(\ref{signal}) by
putting $z=0$. Finally, the outgoing, diffracted field $E_{\ell}$
is calculated from the completely defined r.h.s. of
Eq.\,(\ref{signal}).

\begin{figure}
\includegraphics*[width=8cm]{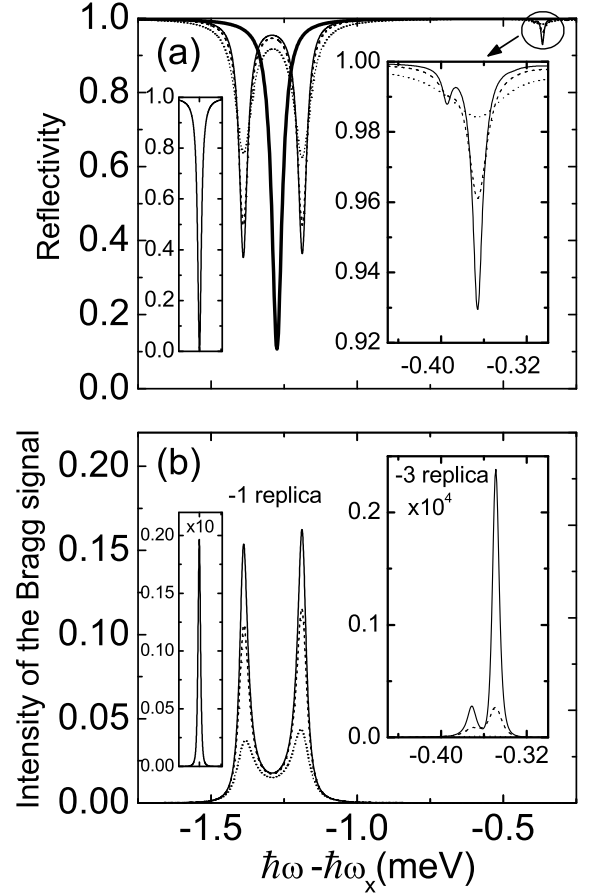}
\caption{ The Bragg spectra of SAW-driven MC polaritons. The
incident optical wave excites the MC polariton states at the
r.h.s. boundary of the acoustically-induced first Brillouin zone
(see Fig.\,1), i.e., $p_{\|}=k/2$ and $\alpha = \alpha_{\rm B}
\simeq 8.2^{\circ}$. (a) The reflection coefficient
$|r(\omega,I_{\rm ac})|^2$ of the optical wave against the
frequency $\omega$. The thick solid line shows the initial
polariton-mediated MC reflectivity for $I_{\rm ac} = 0$ and
$\gamma_{\rm x} = 10\,\mu$eV. The thin lines refer to $|r(I_{\rm
ac}\!\!=\!\!10\,\mbox{W/cm}^2)|^2$ for spectral vicinity of $n=1$
SAW-phonon transition. Inset: $|r|^2$ for spectral vicinity of
$n=3$ SAW-phonon transition. (b) The $-1$ and $-3$ Bragg
diffraction replicas normalized to $I_{\rm opt}$. The l.h.s.
inset: $-1$ Bragg replica calculated by using the nonresonant
$\chi^{(2)}_{\gamma-{\rm ac}}$. The solid, dashed and dotted lines
refer to the excitonic damping $\hbar \gamma_{\rm x} = \hbar/T_2 =
5, 10$, and 30\,$\mu$eV, respectively. }
\label{fig2}
\end{figure}

The Bragg angle $\Theta_{\rm B}$ corresponds to the effective
SAW-induced diffraction of MC polaritons with $p_{\|} = \pm k/2$
and is given by $\sin \Theta_{\rm B} = [k/(2 k_{\rm opt})]\{1 -
[(v_{\rm s}k)/(8 \varepsilon_b \omega_{\rm T})][(ck)/\Omega^{\rm
MC}_{\rm x}]^2\} \simeq k/(2 k_{\rm opt})$, where $k_{\rm opt} =
\omega/c$ is the wavevector of the external optical wave. In
Fig.\,2a we plot the reflection coefficient $|r(\omega,I_{\rm
ac})|^2$ of the incoming light incident at angle $\alpha
=\Theta_{\rm B}$ on the SAW-driven MC ($\nu_{\rm SAW}=1$\,GHz). In
this case the r.h.s. boundary at $p_{\|}=k/2$ of the
acoustically-induced first Brillouin zone is probed (see Fig.\,1).
In the vicinity of $\omega = \omega_{\rm LB}^{\rm MC}(k/2)$, with
increasing $I_{\rm ac}$ the reflectivity changes its single-line
shape for $I_{\rm ac} = 0$ (the initial reflection spectrum is
shown in Fig.\,2a by the bold solid line) to a high contrast
double-line shape with the separation $\Delta^{\rm MC}_{\rm ac}$
between two dips. Furthermore, a similar double-line structure,
which refers to the three-phonon-assisted transition, appears and
develops with increasing $I_{\rm ac}$ for the reflectivity at
$\omega \simeq \omega_{\rm LB}^{\rm MC}(3k/2)$ (see the inset of
Fig.\,2a). Because the wavevector band $-k/2 \leq p_{\|} \leq k/2$
can also be interpreted in terms of the SAW-induced {\it reduced}
Brillouin zone, the incident optical wave probes the odd-order
SAW-induced energy gaps (the areas marked in Fig.\,1 by large
solid and dashed circles refer to the $n=1$ and 3 transitions).
The corresponding frequency-down-shifted $-1$ and $-3$ Bragg
replicas give rise to the outgoing optical signal and are plotted
in Fig.\,2b. The camel-back shape of the replicas follow the band
gaps $\Delta^{\rm MC}_{\rm ac}$ and $\Delta^{\rm MC (n=3)}_{\rm
ac}$, respectively. In Fig.\,3a, $|r(\omega,I_{\rm ac})|^2$ is
shown for $\alpha \simeq 2\Theta_{\rm B}$, so that the r.h.s.
boundary at $p_{\|}=k$ of the second (extended) Brillouin zone is
probed. The corresponding $-2$ Bragg replica is plotted in
Fig.\,3b. In Figs.\,3c-3d we show the energy separation between
two spikes in the $-1$ and $-2$ replicas. The separation is equal
to $\Delta^{\rm MC}_{\rm ac} \propto I_{\rm ac}^{1/2}$ and
$\Delta^{\rm MC (n=2)}_{\rm ac} \propto I_{\rm ac}$, respectively.

\begin{figure}
\includegraphics*[width=8cm]{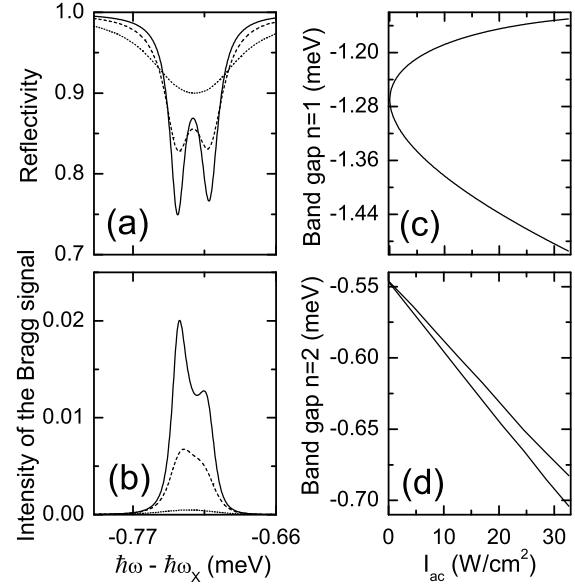}
\caption{ The Bragg spectra of SAW-driven MC polaritons for $n=2$
transition. (a) The reflection coefficient $|r(\omega,I_{\rm
ac})|^2$ for $\alpha = 16.8^{\circ}$ ($p_{\|} \simeq k$ and
$\omega \simeq \omega^{\rm MC}_{\rm LB}(k)$, see Fig.\,1),
$\nu_{\rm SAW} = 1$\,GHz and $I_{\rm ac} = 40$\,W/cm$^2$. (b) The
frequency-down-converted $-2$ Bragg replica. The solid, dashed and
dotted lines refer to the excitonic damping $\hbar \gamma_{\rm x}
= \hbar/T_2 = 5, 10$, and 30\,$\mu$eV, respectively. (c)-(d) The
SAW-induced spike separation, i.e., the band gaps
$\Delta^{\rm MC (n=1)}_{\rm ac}$ and $\Delta^{\rm MC (n=2)}_{\rm ac}$,
against $I_{\rm ac}$. }
\label{fig3}
\end{figure}

Bragg diffraction, due to the nonresonant $\chi^{(2)}_{\gamma-{\rm
ac}}$, has been observed for an optical wave guided by SAW-driven
semiconductor layers \cite{Kuhn70,Loh76,Ciplys02}. The one-phonon
diffraction replica $-1$, calculated for the GaAs-based MC with
the use of $\chi^{(2)}_{\gamma-{\rm ac}}$, is plotted in the
l.h.s. inset of Fig.\,2b. In this case the acoustically-induced
main stop gap in the MC photon spectrum is very small and
completely screened by the radiative damping, so that it cannot be
seen in the Bragg signal as a camel-back structure. Note that the
SAW intensities $I_{\rm ac}$ we discuss are much less than those
used to acoustically ionize, by the piezoelectric effect, the
excitons in GaAs structures \cite{Rocke98,Santos98}.

For the analyzed scattering of MC polaritons by the SAW, the
Klein-Cook parameter \cite{Korpel97} $Q = Q(\omega)$, which
distinguishes the Raman-Nath (transmission-type) regime from the
Bragg (reflection-type) mode, is given by $Q = v_{\rm
pol}(\hbar/\gamma_{\rm pol}) (k^2/k_{\rm opt})$. Here, $v_{\rm
pol}(\omega) $ and $\gamma_{\rm pol}(\omega)$ are the MC polariton
velocity and damping, respectively. For the case analyzed in
Figs.\,1-3, $Q \simeq 6.5 \gg1$, i.e., our theory does deal with
the  Bragg diffraction regime.

We appreciate valuable discussions with Yu. Kosevich, Peter
Littlewood, and W. Sohler. Support of this work by the CREST,
Grant-in-Aid of the Ministry of Education of Japan, Toyota
Phys.\,$\&$\,Chem. Institute, and EU RTN Project HPRN-2002-00298
is gratefully acknowledged.

\end{document}